\PassOptionsToPackage{unicode}{hyperref}
\PassOptionsToPackage{hyphens}{url}
\documentclass[
]{article}
\usepackage{amsmath,amssymb}
\usepackage{iftex}
\ifPDFTeX
  \usepackage[T1]{fontenc}
  \usepackage[utf8]{inputenc}
  \usepackage{textcomp} 
\else 
  \usepackage{unicode-math} 
  \defaultfontfeatures{Scale=MatchLowercase}
  \defaultfontfeatures[\rmfamily]{Ligatures=TeX,Scale=1}
\fi
\usepackage{lmodern}
\ifPDFTeX\else
\fi
\IfFileExists{upquote.sty}{\usepackage{upquote}}{}
\IfFileExists{microtype.sty}{
  \usepackage[]{microtype}
  \UseMicrotypeSet[protrusion]{basicmath} 
}{}
\makeatletter
\@ifundefined{KOMAClassName}{
  \IfFileExists{parskip.sty}{%
    \usepackage{parskip}
  }{
    \setlength{\parindent}{0pt}
    \setlength{\parskip}{6pt plus 2pt minus 1pt}}
}{
  \KOMAoptions{parskip=half}}
\makeatother
\usepackage{xcolor}
\usepackage{longtable,booktabs,array}
\usepackage{calc} 
\usepackage{etoolbox}
\makeatletter
\patchcmd\longtable{\par}{\if@noskipsec\mbox{}\fi\par}{}{}
\makeatother
\IfFileExists{footnotehyper.sty}{\usepackage{footnotehyper}}{\usepackage{footnote}}
\makesavenoteenv{longtable}
\ifLuaTeX
  \usepackage{selnolig}  
\fi
\IfFileExists{bookmark.sty}{\usepackage{bookmark}}{\usepackage{hyperref}}
\IfFileExists{xurl.sty}{\usepackage{xurl}}{} 
\hypersetup{
  pdftitle={Refutable Exclusion Restrictions in Competing Risks with Categorical Covariates},
  pdfauthor={Isfandiyor Akhmedov},
  hidelinks,
  pdfcreator={LaTeX via pandoc}}

\title{Refutable Exclusion Restrictions in Competing Risks with
Categorical Covariates}
\author{Isfandiyor Akhmedov}
\date{August 2026}

\begin{document}
\maketitle

\begin{center}
\small Management Development Institute of Singapore in Tashkent\\
\small \texttt{iakhmedov@mdist.uz}
\end{center}

\hypertarget{abstract}{%
\subsection{Abstract}\label{abstract}}

Competing-risks data do not identify latent marginal duration
distributions or their dependence without additional restrictions. This
paper asks whether restrictions introduced to restore identification
themselves restrict the observable law. For a two-risk Archimedean model
with categorical exclusion restrictions, we derive a
necessary-and-sufficient observable characterization. A discrete
single-crossing argument identifies the scalar copula parameter from
cell-specific overall survival probabilities, while cause indicators
recover the remaining allocation and generate additional specification
restrictions. We construct an identification-robust, self-normalized
quadratic statistic, invert it to obtain confidence sets, and use empty
inverted sets as a conservative specification test. Under complete
competing-risks observation, simulations show that the cause indicator
can be decisive: under the weakest contrast considered it converts a
frequently uninformative survival-based confidence set into an
informative joint set without loss of coverage, whereas excessive
categorical contrast can eliminate causes from individual cells and make
cause-specific recovery inadmissible. The results turn latent exclusion
restrictions into refutable restrictions without estimating covariate
derivatives.

\textbf{Keywords:} competing risks; exclusion restrictions;
refutability; Archimedean copula; categorical covariates; weak
identification.

\hypertarget{introduction}{%
\section{1. Introduction}\label{introduction}}

Competing-risks data reveal the time and type of the first event but not
the latent times at which the remaining events would have occurred. This
loss of information is structural. As shown by Tsiatis (1975), distinct
joint laws of the latent durations can generate exactly the same
distribution of the observed minimum and its cause. Consequently,
neither the latent marginal distributions nor their dependence are
identified from the observed competing-risks law without additional
restrictions. This result is often read as a reason to choose an
identifying model and proceed conditionally on it. A large subsequent
literature, including Heckman and Honoré (1989), restores identification
by imposing additional structure on dependence, heterogeneity, or
covariate variation. A different question is whether the identifying
restrictions themselves leave observable implications. In other words,
once a restriction is introduced to overcome nonidentification, can the
data ever refute it?

This paper answers that question for a competing-risks model with
Archimedean dependence and categorical exclusion restrictions. Each of
two covariates is excluded from one latent marginal duration while being
allowed to shift the other. The restrictions are imposed on latent
marginals and therefore cannot be verified by directly observing the
excluded outcome. Nevertheless, they are not observationally vacuous.
They imply a finite set of restrictions on cell-specific observable
functionals. These restrictions identify the copula dependence
parameter, recover the latent marginal survival functions over the time
interval supported by the data, and generate a specification test. Thus
an assumption introduced to restore identification also restricts the
observable law that it is meant to rationalize.

The distinction between identification and refutability is central. Let
\(\Psi\) map a latent competing-risks law into the law of the observed
duration and cause, and let \(\mathcal S\) denote the maintained model
restrictions. Identification asks whether a parameter or latent
functional is constant on the fibers of \(\Psi\) within \(\mathcal S\).
Refutability asks whether the observable image \(\Psi(\mathcal S)\) is a
strict subset of the unrestricted set of observable laws. Neither
property implies the other. A restriction may select a unique latent
completion for every observable law and hence identify the model without
being refutable; conversely, it may rule out some observable laws while
leaving multiple latent completions for those that remain. Our results
establish both properties separately. The copula parameter is identified
by an observable cross-difference, while the cause indicators resolve
the remaining allocation freedom and supply additional specification
restrictions.

The starting point is the overall conditional survival probability
\(\pi(t,z)=\Pr(T>t\mid Z=z)\). Under an Archimedean copula with
generator \(\phi_\theta\) and the two exclusion restrictions, the
transformed survival \(\phi_\theta\{\pi(t,z_1,z_2)\}\) must be
additively separable in \(z_1\) and \(z_2\). For binary covariates this
is equivalent, at every time point, to the vanishing cross-difference

\[
\Delta_t(\theta)=
\phi_\theta(\pi_{00}(t))+\phi_\theta(\pi_{11}(t))
-\phi_\theta(\pi_{01}(t))-\phi_\theta(\pi_{10}(t)).
\]

Under the standard ordering condition on Archimedean generators, the
sign of \(\Delta_t(\theta)\) changes only once. The proof is a discrete
single-crossing argument: after composing a candidate generator with the
inverse true generator, convexity or concavity gives the sign of the
mixed finite difference. Hence one informative \(2\times2\) subtable at
one time point identifies \(\theta\); additional time points and larger
categorical tables provide overidentifying restrictions. The strength of
this identification is governed by the product of the two cell
contrasts. This yields an observable design-strength diagnostic and
explains why theoretical identification can be weak in samples when
either excluded covariate has little effect.

The first-stage cross-difference uses only the overall survival law. The
cause indicator has a different and precisely characterizable role.
Define the observable cause-specific transform

\[
A_{j,\theta}(t,z)=
E\!\left[-\phi_\theta'\{\pi(T,z)\}
\mathbf 1\{T\le t,J=j\}\mid Z=z\right].
\]

For every candidate parameter and every observable law,
\(A_{1,\theta}+A_{2,\theta}=\phi_\theta(\pi)\). The sum therefore
contains no cause-specific identifying information: it is an identity
induced by the conditional probability integral transform. Once the
first-stage additive decomposition is imposed, the two cause-specific
restrictions become two projections of a single residual allocation
function. Jointly, they require that function to equal a time-specific
common normalization \(c(t)\). The cause indicator thus fixes exactly
the freedom left by the nonunique additive decomposition. This recovers
the two latent marginal survival functions and creates specification
restrictions beyond those used to identify \(\theta\).

These results relate most closely to Hiabu, Lo, and Wilke (2025), who
establish identification under exclusion restrictions and Archimedean
dependence using cross-derivatives of the transformed overall survival
function. Their argument requires continuously distributed covariates,
twice differentiable conditional survival, and nonvanishing covariate
derivatives (Assumption 1(ii) in the 2023 preprint version); their
estimator correspondingly relies on kernel estimates of
cross-derivatives. Their identification theorem and estimation procedure
therefore do not cover categorical designs. The present paper replaces
derivatives by finite cross-differences, proves identification for
categorical designs, and develops finite-dimensional inference based on
cell means. The discrete single-crossing proof parallels the ordering
logic behind the continuous identification result, so the
finite-difference lemma is not presented as an unrelated identification
principle. The substantive additions are the observable
characterization, recovery of the latent allocation, and an
identification-robust specification procedure for a setting in which the
derivative-based estimator is not defined.

The paper also connects to the broader literature on the empirical
content of identifying assumptions, including sharp characterizations
and testability in incomplete structural models (Mourifié, Henry, and
Méango, 2020). Liu (2020) provides a complementary
structural-identification analysis for competing risks with time-varying
heterogeneity and simultaneous failure; it is not a predecessor of the
specification test developed here. That literature asks whether
restrictions introduced into partially or nonidentified models merely
select a latent representation or instead carve out a proper subset of
observable distributions. We provide such a characterization for an
Archimedean competing-risks model. The result is deliberately
model-specific: rejection means that the observed law is incompatible
with the maintained combination of the copula family, exclusion
restrictions, and stability of the dependence parameter. It does not
identify which component fails. A natural sensitivity analysis therefore
reports compatibility across a finite collection of copula families
rather than interpreting rejection of one family as rejection of the
exclusions alone. Near independence those families become intrinsically
difficult to distinguish, which is a feature of the statistical problem
rather than a defect of the test.

The setting is practically relevant whenever only the first of several
mutually exclusive exits is observed and quantitative conclusions depend
on their latent association. Examples include different exits from
unemployment, disease progression and death, and claim occurrence and
policy lapse in insurance portfolios. In such applications the most
credible excluded variables are often categorical institutional
features--contract options, administrative assignments, benefit rules,
or treatment pathways--rather than continuously varying instruments. The
exclusions must still be defended conditionally on common observed
covariates; the value of a specification test is precisely that this
substantive argument no longer has to be accepted without observable
implications.

Inference is based on a continuously updated, self-normalized quadratic
statistic in the spirit of Hansen, Heaton, and Yaron (1996). At each
candidate \(\theta\), the moment covariance is recomputed and the
statistic is inverted to obtain a confidence set. This construction is
important in the present model because the natural Clayton
transformations can change scale exponentially with \(\theta\). A
fixed-weight minimum-distance criterion is pointwise consistent under
standard separation conditions, but it can behave catastrophically in
finite samples because it does not absorb this parameter-dependent
scale. Continuously updated normalization does. We use the statistic at
a fixed candidate value for identification-robust inference: under the
null its reference distribution does not rely on the estimator being
approximately normal or on strong local identification. Empty inverted
sets provide a conservative specification test: the rejection event
based on the infimum is contained in the rejection event at the true
parameter. This event-inclusion argument is distinct from the limiting
distribution of the minimized criterion. We do not treat that minimum as
a chi-square overidentification statistic under weak identification,
because its limit is generally an infimum of a Gaussian process rather
than a standard chi-square law (Stock and Wright, 2000).

Cause-specific moments introduce a second practical issue. Although
finite categorical cells eliminate the divergent covariate-tail moments
encountered in unbounded continuous designs, the weights can still
become highly uneven at late event times. A few observations may then
dominate the covariance estimate and the quadratic statistic. We
therefore treat the usable time horizon as part of the inferential
output. From a finite, prespecified family of time grids, the procedure
selects the latest grid satisfying minimum-exposure, full-rank,
fourth-moment, and largest-observation diagnostics uniformly over the
survival-block confidence set. If no grid is admissible, cause-specific
inference is withheld while the survival-block confidence set remains
reportable. The accompanying diagnostics distinguish a lack of
worst-cell exposure from domination of the influence-function covariance
by a small number of observations.

The paper makes five contributions. First, it establishes identification
of Archimedean dependence from categorical exclusions through a discrete
single-crossing condition and provides a direct measure of design
strength. Second, it gives a necessary-and-sufficient observable
characterization of the restricted model and proves strict refutability:
the cause-specific model is a proper subset of the observable laws
satisfying the first-stage cross-difference restrictions alone. Third,
it characterizes the exact contribution of the observed cause: recovery
of the residual allocation \(c(t)\), the latent marginal survival
functions, and additional testable restrictions. Fourth, it constructs a
joint self-normalized statistic whose inversion delivers confidence sets
and a conservative specification test without requiring strong
identification. Fifth, it gives an operational rule for determining the
time horizon over which cause-specific inference is supported by the
data, accompanied by diagnostics that distinguish inadequate cell
exposure from domination by extreme influence values.

The rest of the paper is organized as follows. Section 2 separates
identification from refutability, derives the cross-difference
condition, and characterizes the role of the cause indicator. Section 3
presents estimation, joint inference, grid admissibility, and the
horizon-selection rule. Section 4 reports the simulation evidence,
including weak and strong categorical designs and the finite-sample
behavior of the survival and cause-specific blocks. Section 5 discusses
extensions, including right censoring and sensitivity to the maintained
copula family. Appendix A gives the omitted proofs.

\hypertarget{observable-content-of-categorical-exclusions}{%
\section{2. Observable content of categorical
exclusions}\label{observable-content-of-categorical-exclusions}}

\hypertarget{identification-and-refutability}{%
\subsection{2.1 Identification and
refutability}\label{identification-and-refutability}}

Let \(T_1,T_2\) be latent event times, \(T=\min(T_1,T_2)\), and
\(J\in\{1,2\}\) the observed cause. Let \(Z=(Z_1,Z_2)\), with
\(Z_1,Z_2\in\{0,1\}\), and condition throughout on any common
categorical covariates \(X=x\). The map from a latent joint law to the
law of \((T,J,Z)\) is denoted by \(\Psi\). A scalar parameter is
identified on a restricted model \(\mathcal S\) when it is constant on
every fiber of \(\Psi\) intersected with \(\mathcal S\). The full latent
law is identified when \(\Psi|_{\mathcal S}\) is injective. Refutability
is different: \(\mathcal S\) is refutable when \(\Psi(\mathcal S)\) is a
strict subset of the unrestricted observable laws.

\hypertarget{archimedean-model-and-exclusions}{%
\subsection{2.2 Archimedean model and
exclusions}\label{archimedean-model-and-exclusions}}

Write

\[
S_j(t\mid z)=\Pr(T_j>t\mid Z=z),\quad
\pi(t,z)=\Pr(T>t\mid Z=z).
\]

For a one-parameter Archimedean survival copula with decreasing
generator \(\phi_\theta\),

\[
\phi_\theta\{\pi(t,z)\}
=\phi_\theta\{S_1(t\mid z)\}+\phi_\theta\{S_2(t\mid z)\}.
\tag{2.1}\label{eq:archimedean}
\]

The categorical exclusion restrictions are

\[
S_1(t\mid z_1,z_2)=S_1(t\mid z_1),\quad
S_2(t\mid z_1,z_2)=S_2(t\mid z_2).
\tag{2.2}\label{eq:exclusions}
\]

Thus the transformed observable survival must be additively separable:

\[
\phi_\theta\{\pi_{ab}(t)\}=a_a(t)+b_b(t),\qquad a,b\in\{0,1\}.
\tag{2.3}\label{eq:additive}
\]

Equivalently,

\[
\Delta_t(\theta)=
\phi_\theta(\pi_{00})+\phi_\theta(\pi_{11})
-\phi_\theta(\pi_{01})-\phi_\theta(\pi_{10})=0.
\tag{2.4}\label{eq:cross-difference}
\]

\textbf{Assumption 1 (regular Archimedean family and observable
support).} The parameter set is compact. On a common interior
probability domain containing all retained cell survivals, each
\(\phi_\theta\) is convex, twice continuously differentiable, and
invertible, with \(\phi_\theta'(u)<0\), the normalization
\(\phi_\theta(1)=0\), and a common admissible range for transformed
sums. Every retained cell has positive probability, and the observable
survival and cause subdistributions are continuous on the retained time
interval.

The Archimedean representation and the exclusion restrictions are not
auxiliary regularity assumptions: together they define the refutable
null model \(\mathcal S\). Assumption 1 supplies the regularity and
common-domain conditions under which that null is characterized.

\hypertarget{discrete-single-crossing}{%
\subsection{2.3 Discrete single
crossing}\label{discrete-single-crossing}}

Assume the observable law is generated at \(\theta_0\) and, for some
\((t,x)\), both transformed cell contrasts are nonzero. For
\(\theta\neq\theta_0\), define

\[
H_{\theta,\theta_0}(u)=\phi_\theta\{\phi_{\theta_0}^{-1}(u)\}.
\]

\textbf{Assumption 2 (ordered generators).} For every
\(\theta_1>\theta_2\), the ratio
\(\phi'_{\theta_2}(s)/\phi'_{\theta_1}(s)\) is strictly increasing in
\(s\) on the common interior domain.

Under Assumption 2, \(H_{\theta,\theta_0}\) is strictly convex for
\(\theta>\theta_0\), strictly concave for \(\theta<\theta_0\), and the
identity at \(\theta=\theta_0\). If
\(\phi_{\theta_0}(\pi_{ab})=\alpha_a+\beta_b\), signed integration gives

\[
\Delta_t(\theta)=\int_{\alpha_0}^{\alpha_1}
\{H'_{\theta,\theta_0}(u+\beta_1)
-H'_{\theta,\theta_0}(u+\beta_0)\}\,du.
\tag{2.5}\label{eq:mixed-integral}
\]

Hence

\[
\operatorname{sign}\Delta_t(\theta)=
\operatorname{sign}\{(\alpha_1-\alpha_0)(\beta_1-\beta_0)(\theta-\theta_0)\}.
\tag{2.6}\label{eq:single-crossing-sign}
\]

\textbf{Proposition 1 (categorical identification).} Under the
common-domain and generator-ordering conditions, one \(2\times2\) table
at one time point for which both transformed cell contrasts are nonzero
identifies \(\theta_0\) uniquely from overall survival. Additional times
and strata supply overidentifying restrictions.

This result replaces the nonzero cross-derivative condition used for
continuous covariates by nonzero finite contrasts. It proves uniqueness,
not uniform finite-sample strength.

\hypertarget{cause-specific-observable-transforms}{%
\subsection{2.4 Cause-specific observable
transforms}\label{cause-specific-observable-transforms}}

Define, for \(j=1,2\),

\[
A_{j,\theta,c}(t)=E\!\left[-\phi_\theta'\{\pi_c(T)\}
\mathbf 1\{T\le t,J=j\}\mid C=c\right],
\tag{2.7}\label{eq:cause-transform}
\]

where \(c=(a,b,x)\). At the true value,

\[
A_{1,\theta_0,abx}(t)=\phi_{\theta_0}\{S_1(t\mid a,x)\},\quad
A_{2,\theta_0,abx}(t)=\phi_{\theta_0}\{S_2(t\mid b,x)\}.
\tag{2.8}\label{eq:copula-graphic}
\]

The identity in \eqref{eq:copula-graphic} is the Archimedean
copula-graphic representation of Zheng and Klein (1995) and Rivest and
Wells (2001). Our contribution is the exclusion-restricted compatibility
condition that this representation implies across categorical cells.

Moreover, for every candidate \(\theta\) and every observable law,

\[
A_{1,\theta,c}(t)+A_{2,\theta,c}(t)=\phi_\theta\{\pi_c(t)\}.
\tag{2.9}\label{eq:cause-sum}
\]

This follows from the conditional probability integral transform; under
the continuity condition in Assumption 1, \(\pi_c(T)\mid C=c\) is
uniform on \((0,1)\). The same continuity condition permits the
Stieltjes change of variables used in Lemma A.0, so the two appearances
of continuity are not separate assumptions. Both transforms are
estimable from cell averages and require neither cause-specific density
estimation nor covariate smoothing.

Suppose \eqref{eq:cross-difference} holds and choose any decomposition
\(\phi_\theta(\pi_{abx})=a_{a,x}+b_{b,x}\). Let

\[
R_{ab,x}(t)=A_{1,\theta,abx}(t)-a_{a,x}(t).
\tag{2.10}\label{eq:residual}
\]

Then \eqref{eq:cause-sum} implies \(A_{2,\theta,abx}=b_{b,x}-R_{ab,x}\).
On a full positive-probability table, both exclusions hold if and only
if

\[
R_{ab,x}(t)=c_x(t)\qquad\text{for all }a,b.
\tag{2.11}\label{eq:compatibility}
\]

The observed cause therefore fixes the time-specific normalization left
unidentified by additive decomposition.

\hypertarget{observable-characterization}{%
\subsection{2.5 Observable
characterization}\label{observable-characterization}}

\textbf{Theorem 1 (necessary and sufficient characterization).} Fix a
candidate \(\theta\). An observable competing-risks law is
rationalizable by the Archimedean model and the categorical exclusions
on an interval \([0,\bar t)\) if and only if: (i) the overall-survival
cross-differences \eqref{eq:cross-difference} vanish for all retained
times and strata; (ii) the cause-specific residuals satisfy
\eqref{eq:compatibility}; and (iii) the constructed transformed marginal
cumulative hazards are nondecreasing from zero and lie in the domain of
\(\phi_\theta^{-1}\). Here rationalizability on \([0,\bar t)\) means
that the generated law of \((T,J)\) agrees with the observable law on
that interval; the latent marginals may be continued beyond \(\bar t\)
in any manner compatible with valid survival functions and the chosen
copula. When these conditions hold, the latent marginals are recovered
constructively as

\[
S_1(t\mid a,x)=\phi_\theta^{-1}\{a_{a,x}(t)+c_x(t)\},\quad
S_2(t\mid b,x)=\phi_\theta^{-1}\{b_{b,x}(t)-c_x(t)\}.
\tag{2.12}\label{eq:recovered-marginals}
\]

Necessity follows from \eqref{eq:archimedean}, \eqref{eq:exclusions},
and \eqref{eq:copula-graphic}. For sufficiency,
\eqref{eq:recovered-marginals} constructs valid marginal survival
functions; their Archimedean combination reproduces \(\pi\), while
\eqref{eq:copula-graphic} reproduces the observed cause allocation.

\hypertarget{strict-refutability}{%
\subsection{2.6 Strict refutability}\label{strict-refutability}}

The cause-specific conditions are not consequences of the survival
cross-difference. Start from any law satisfying the model at
\(\theta_0\) with informative contrasts. Perturb the two cause-specific
subdistributions by equal and opposite small functions, preserving their
sum and hence preserving \(\pi\) and \(\Delta_t(\theta_0)=0\), but
choose the perturbation so that \eqref{eq:compatibility} fails.
Proposition 1 makes \(\theta_0\) the only candidate satisfying the
first-stage restriction; at that candidate the perturbed cause
allocation violates the exclusions. Therefore no \(\theta\) rationalizes
the perturbed law.

\textbf{Corollary 1.} The observable model satisfying all categorical
exclusions is a strict subset of the observable laws satisfying the
overall-survival cross-difference alone.

At a fixed time the overall-survival block contributes one independent
contrast, while the four cause-allocation cells contribute three
independent residual contrasts after using \eqref{eq:cause-sum}. This
rank count is useful bookkeeping, but identification and strict
inclusion follow from the constructive and perturbation arguments above,
not from dimension counting alone.

\hypertarget{statistical-procedure}{%
\section{3. Statistical procedure}\label{statistical-procedure}}

\hypertarget{survival-block}{%
\subsection{3.1 Survival block}\label{survival-block}}

Choose prespecified time nodes \(t_1,\ldots,t_q\). For cell \(c\),
estimate

\[
\widehat\pi_c(t)=n_c^{-1}\sum_{i:C_i=c}\mathbf1\{T_i>t\}.
\]

The influence function under whole-sample normalization is

\[
IF_{\pi_c(t)}(O_i)=\frac{\mathbf1\{C_i=c\}}{p_c}
\{\mathbf1(T_i>t)-\pi_c(t)\}.
\tag{3.1}\label{eq:if-survival}
\]

Let \(g_{\pi,n}(\theta)\) stack the empirical cross-differences over
nodes and strata. Its covariance is estimated from the stacked
individual influence vectors, including the derivative of
\(\phi_\theta\). The survival statistic is

\[
\mathcal A_{\pi,n}(\theta)=n\,g_{\pi,n}(\theta)'
\widehat\Omega_\pi(\theta)^+g_{\pi,n}(\theta).
\tag{3.2}\label{eq:survival-statistic}
\]

The CUE estimate minimizes \eqref{eq:survival-statistic}, while the
primary inferential object is its inversion.

\hypertarget{cause-specific-block-and-plug-in-influence}{%
\subsection{3.2 Cause-specific block and plug-in
influence}\label{cause-specific-block-and-plug-in-influence}}

With \(w_\theta(u)=-\phi_\theta'(u)\), estimate
\eqref{eq:cause-transform} by

\[
\widehat A_{1,\theta,c}(t)=n_c^{-1}\sum_{i:C_i=c}
w_\theta\{\widehat\pi_c(T_i)\}\mathbf1\{T_i\le t,J_i=1\}.
\tag{3.3}\label{eq:cause-estimator}
\]

Because the weight contains the estimated survival function,
\eqref{eq:cause-estimator} is a plug-in statistic. Its within-cell
influence function is

\[
IF_{A_1}(O_i;t,c)=w_\theta\{\pi_c(T_i)\}
\mathbf1\{T_i\le t,J_i=1\}-A_{1,\theta,c}(t)+B_{1,\theta,c}(O_i;t),
\tag{3.4}\label{eq:if-cause}
\]

where

\[
B_{1,\theta,c}(O_i;t)=\int_0^t w_\theta'\{\pi_c(s)\}
\{\mathbf1(T_i>s)-\pi_c(s)\}\,dF_{1,c}(s).
\]

Multiplication by \(\mathbf1(C_i=c)/p_c\) converts this to whole-sample
normalization. The analogous expression holds for cause 2. Omitting the
second term understates uncertainty.

Let \(D\) denote the design matrix for the nuisance additive components
and choose a fixed full-row-rank contrast matrix \(Q\) satisfying
\(QD=0\). Applying \(Q\) to the stacked cause-specific cell transforms
removes the nuisance decomposition without data-dependent GLS
circularity. Any nonsingular change of basis \(Q_2=BQ_1\) leaves the
population quadratic form exactly invariant when the covariance on the
contrast space is nonsingular.

\hypertarget{joint-statistic-and-rank}{%
\subsection{3.3 Joint statistic and
rank}\label{joint-statistic-and-rank}}

Stack survival and cause-specific moments into \(g_n(\theta)\). For each
observation assemble one joint influence vector, including every
cross-block covariance, and set

\[
\widehat\Omega(\theta)=n^{-1}\sum_{i=1}^n
\widehat{IF}_i(\theta)\widehat{IF}_i(\theta)'.
\tag{3.5}\label{eq:joint-covariance}
\]

The empirical influences are centered by construction. Componentwise
scaling is used for numerical conditioning but leaves the quadratic form
unchanged. The joint statistic is

\[
\mathcal A_n(\theta)=n\,g_n(\theta)'\widehat\Omega(\theta)^+g_n(\theta).
\tag{3.6}\label{eq:joint-statistic}
\]

Theoretical rank \(r_0\) is distinguished from a numerical rank
estimate. A prespecified tolerance is used only to diagnose whether the
intended moment system retains full rank; rank loss makes a candidate
grid inadmissible rather than silently lowering degrees of freedom.

\hypertarget{confidence-sets-and-specification-testing}{%
\subsection{3.4 Confidence sets and specification
testing}\label{confidence-sets-and-specification-testing}}

Let \(c_{1-\alpha,n}(\theta)\) be the prespecified critical value.
Define

\[
\widehat C_{1-\alpha}=
\{\theta\in\Theta_0:\mathcal A_n(\theta)\le c_{1-\alpha,n}(\theta)\}.
\tag{3.7}\label{eq:confidence-set}
\]

For the regular survival block,
\(c_{1-\alpha,n}=\chi^2_{r_0,1-\alpha}\). Cause-specific directions can
have heavy finite-sample tails. The working calibration uses

\[
c^F_{1-\alpha,n}(\theta)=
\frac{r_0\{\widehat m_{\mathrm{eff}}(\theta)-1\}}
{\widehat m_{\mathrm{eff}}(\theta)-r_0}
F_{r_0,\widehat m_{\mathrm{eff}}(\theta)-r_0;1-\alpha},
\tag{3.8}\label{eq:f-critical}
\]

only when \(\widehat m_{\mathrm{eff}}>r_0\). Noninteger degrees of
freedom are defined through the beta-function representation. The
effective size is a robust working calibration based on directional
normalized fourth moments, not a derived exact pivot; therefore it is
used jointly with tail diagnostics rather than alone.

An empty inverted set rejects the maintained model. If \(\theta_0\)
belongs to the null, then

\[
\{\inf_{\theta\in\Theta_0}[\mathcal A_n(\theta)-c_{1-\alpha,n}(\theta)]>0\}
\subseteq
\{\mathcal A_n(\theta_0)>c_{1-\alpha,n}(\theta_0)\}.
\tag{3.9}\label{eq:event-inclusion}
\]

Consequently the asymptotic rejection probability is at most \(\alpha\);
equality is not claimed. This projection argument does not assign a
chi-square law to the minimized criterion.

\hypertarget{strength-and-tail-diagnostics}{%
\subsection{3.5 Strength and tail
diagnostics}\label{strength-and-tail-diagnostics}}

The survival-block design-strength diagnostic is

\[
\widehat{\mathcal S}=n\,G(\widehat\theta)'
\widehat\Omega_\pi(\widehat\theta)^+G(\widehat\theta),
\tag{3.10}\label{eq:strength}
\]

where \(G\) is the derivative of the survival moment vector. It plays
the role of a first-stage strength measure, but depends on the chosen
time grid and should not be compared across different systems of
moments.

For the cause-specific block, report the median, 90th percentile, and
maximum directional fourth-moment index, the estimated effective size,
the largest single-observation share of the empirical second moment,
worst-cell survival and risk-set exposure, and full-rank status. The
median effective-size calibration targets the working 95\% level;
extreme directional indices explain why it need not reproduce the 99\%
tail.

\hypertarget{prespecified-horizon-selection}{%
\subsection{3.6 Prespecified horizon
selection}\label{prespecified-horizon-selection}}

Let \(\mathfrak T=\{\mathcal T^{(1)},\ldots,\mathcal T^{(M)}\}\) be a
finite ordered family of candidate time grids fixed before confirmatory
analysis. Diagnostics are required uniformly over
\(C_{\pi,n}\cap\Theta_0\), where \(C_{\pi,n}\) is the survival-block
confidence set and \(\Theta_0\) is the maintained substantive parameter
range. Select the latest grid satisfying all exposure, rank,
fourth-moment, and largest-observation requirements. If none is
admissible, withhold cause-specific inference and report only the
survival block.

\textbf{Assumption 3 (stable finite-grid selection).} The candidate
family \(\mathfrak T\) is finite and fixed before confirmatory analysis.
On every candidate grid, retained weights are uniformly bounded and the
moments required for consistency of the exposure, rank, fourth-moment,
and largest-observation diagnostics are finite. Their population
counterparts are separated from the prespecified admissibility
thresholds uniformly over the diagnostic parameter region.

Because the family is finite, Assumption 3 gives uniform convergence of
the diagnostic estimates over its members and makes the selected grid
asymptotically deterministic. At threshold boundaries the fixed-grid
coverage statement does not automatically extend to adaptive selection;
sample splitting remains a conservative alternative.

\textbf{Assumption 4 (maintained dependence range).} Before observing
specification-test outcomes, the researcher fixes a compact,
substantively defensible range \(\mathcal T_0\) in Kendall's \(\tau\)
scale and assumes \(\tau_0\in\mathcal T_0\). Each candidate copula
family is restricted to the intersection of \(\mathcal T_0\) with the
Kendall range that the family can represent.

Coverage is conditional on Assumption 4. Contact with
\(\partial\mathcal T_0\) is assumption censoring, while contact with the
wider computational grid is a numerical defect; the two are reported
separately. Theorem 1 uses Assumptions 1--2 but not Assumption 3;
Assumption 3 is needed only for the asymptotic justification of
data-dependent grid selection, and Assumption 4 conditions the reported
confidence statement.

\hypertarget{simulation-evidence}{%
\section{4. Simulation evidence}\label{simulation-evidence}}

\hypertarget{design}{%
\subsection{4.1 Design}\label{design}}

The data-generating process has two binary excluded covariates and four
equally likely cells. Clayton dependence is indexed by Kendall's
\(\tau\in\{0.2,0.5\}\); the marginal covariate contrast is
\(\beta\in\{1,2,3\}\); and \(n\in\{2000,8000\}\). Each configuration
uses 500 independent replications. Time grids are fixed from population
quantiles. For proportions based on 500 replications, the maximum
binomial Monte Carlo standard error is 0.022; for the 2,000-replication
calibration experiment in Section 4.3, it is 0.011.

The survival block uses four cross-difference moments. The joint block
has theoretical rank 16. Confidence sets are obtained by inversion on a
common maintained Kendall range rather than by Wald approximation to the
CUE estimate.

\hypertarget{survival-block-performance}{%
\subsection{4.2 Survival-block
performance}\label{survival-block-performance}}

The survival statistic is well calibrated in informative designs. At
\((\tau_0,\beta,n)=(0.2,2,8000)\), coverage is 0.978 and the empirical
90\%, 95\%, and 99\% quantiles are 7.18, 8.42, and 10.46, against 7.78,
9.49, and 13.28 for \(\chi^2_4\). The difference is conservative and
concentrated in the upper tail; a PIT Kolmogorov-Smirnov test alone does
not reveal it. At \((0.2,3,8000)\), coverage is 0.954 and the 95\% and
99\% quantiles, 9.08 and 12.73, are close to their references.

Weak configurations reveal why coverage cannot be read without set size.
At \(n=2000\) and \(\tau_0=0.5\), the inverted set often occupies most
of the maintained range and reaches its upper boundary. High coverage
there is largely uninformative. The loss of sensitivity is asymmetric:
the cross-difference flattens as positive dependence increases, so upper
endpoints are much less data-determined than lower endpoints. Results
are therefore interpreted primarily in Kendall's \(\tau\) scale and
boundary contacts are reported as assumption-censored.

\textbf{Table 1. Survival-block estimation, confidence-set size,
boundary contact, and design strength.}

\begin{longtable}[]{@{}
  >{\raggedleft\arraybackslash}p{(\columnwidth - 14\tabcolsep) * \real{0.1250}}
  >{\raggedleft\arraybackslash}p{(\columnwidth - 14\tabcolsep) * \real{0.1250}}
  >{\raggedleft\arraybackslash}p{(\columnwidth - 14\tabcolsep) * \real{0.1250}}
  >{\raggedleft\arraybackslash}p{(\columnwidth - 14\tabcolsep) * \real{0.1250}}
  >{\raggedleft\arraybackslash}p{(\columnwidth - 14\tabcolsep) * \real{0.1250}}
  >{\raggedleft\arraybackslash}p{(\columnwidth - 14\tabcolsep) * \real{0.1250}}
  >{\raggedleft\arraybackslash}p{(\columnwidth - 14\tabcolsep) * \real{0.1250}}
  >{\raggedleft\arraybackslash}p{(\columnwidth - 14\tabcolsep) * \real{0.1250}}@{}}
\toprule\noalign{}
\begin{minipage}[b]{\linewidth}\raggedleft
\(\tau_0\)
\end{minipage} & \begin{minipage}[b]{\linewidth}\raggedleft
\(\beta\)
\end{minipage} & \begin{minipage}[b]{\linewidth}\raggedleft
\(n\)
\end{minipage} & \begin{minipage}[b]{\linewidth}\raggedleft
RMSE
\end{minipage} & \begin{minipage}[b]{\linewidth}\raggedleft
Cov.
\end{minipage} & \begin{minipage}[b]{\linewidth}\raggedleft
Mean fraction
\end{minipage} & \begin{minipage}[b]{\linewidth}\raggedleft
Bound.
\end{minipage} & \begin{minipage}[b]{\linewidth}\raggedleft
Median \(\widehat{\mathcal S}\)
\end{minipage} \\
\midrule\noalign{}
\endhead
\bottomrule\noalign{}
\endlastfoot
0.2 & 1 & 2000 & 0.275 & 0.958 & 0.773 & 0.986 & 5.72 \\
0.2 & 1 & 8000 & 0.121 & 0.950 & 0.440 & 0.358 & 11.19 \\
0.2 & 2 & 2000 & 0.162 & 0.966 & 0.415 & 0.926 & 14.85 \\
0.2 & 2 & 8000 & 0.054 & 0.978 & 0.207 & 0.016 & 41.21 \\
0.2 & 3 & 2000 & 0.139 & 0.956 & 0.383 & 0.868 & 15.99 \\
0.2 & 3 & 8000 & 0.044 & 0.954 & 0.168 & 0.000 & 54.86 \\
0.5 & 1 & 2000 & 0.294 & 0.982 & 0.723 & 1.000 & 4.36 \\
0.5 & 1 & 8000 & 0.164 & 0.948 & 0.402 & 0.868 & 5.48 \\
0.5 & 2 & 2000 & 0.215 & 0.990 & 0.396 & 1.000 & 7.80 \\
0.5 & 2 & 8000 & 0.104 & 0.958 & 0.265 & 0.618 & 9.32 \\
0.5 & 3 & 2000 & 0.207 & 0.980 & 0.356 & 1.000 & 5.86 \\
0.5 & 3 & 8000 & 0.084 & 0.960 & 0.239 & 0.456 & 8.44 \\
\end{longtable}

The mean range fraction is the mean length of the inverted survival set
divided by the maintained Kendall-\(\tau\) range. Boundary contact
records intersection with \(\partial\mathcal T_0\). The table makes the
contrast effect visible: at \(\tau_0=0.2\) and \(n=8000\), increasing
\(\beta\) from 1 to 3 reduces RMSE from 0.121 to 0.044, reduces the mean
range fraction from 0.440 to 0.168, eliminates boundary contact, and
raises median \(\widehat{\mathcal S}\) from 11.19 to 54.86.

\hypertarget{joint-calibration-and-effective-size}{%
\subsection{4.3 Joint calibration and effective
size}\label{joint-calibration-and-effective-size}}

The full plug-in influence function is validated by empirical centering
below \(3.5\times10^{-14}\), stable rank 16, and recovery of nominal
calibration as \(n\) increases. At \((0.2,2,8000)\), the uncorrected
joint statistic has 0.933 coverage over 2000 replications. Its median
effective size is 412 (10th--90th percentile 373--452). The working
Hotelling--\(F\) calibration raises coverage to 0.951.
Empirical-to-reference quantile ratios are close to one through the
working 95\% level, while the 99\% tail remains about 5.6\% heavier.
This limitation is consistent with extreme directional fourth moments
that the median calibration intentionally does not track.

The time grid matters more than merely increasing the critical value.
For \(\tau_0=0.5\), \(\beta=2\), and \(n=8000\), the standard late grid
produces a largest-observation share of 0.696 and only 0.362 chi-square
coverage (0.398 after the \(F\) correction). Moving to an earlier grid
reduces the share to 0.176 and restores coverage to 0.962 (0.976 after
correction). For \(\tau_0=0.2\), \(\beta=3\), an early grid similarly
improves chi-square coverage from 0.902 to 0.954. Thus horizon selection
precedes finite-sample calibration.

\hypertarget{confirmatory-horizon-selection-experiment}{%
\subsection{4.4 Confirmatory horizon-selection
experiment}\label{confirmatory-horizon-selection-experiment}}

Table 2a summarizes the prespecified six-grid procedure together with
its prespecified family-adequacy extension. `Admissible' means that at
least one cause-specific grid passed all diagnostics uniformly over the
survival confidence set. Conditional coverage and empty-set frequencies
use only admissible replications; zero denominators are shown as
unavailable rather than as zero.

\textbf{Table 2a. Confirmatory admissibility, coverage, and
specification-test results.}

\begin{longtable}[]{@{}
  >{\raggedleft\arraybackslash}p{(\columnwidth - 14\tabcolsep) * \real{0.1250}}
  >{\raggedleft\arraybackslash}p{(\columnwidth - 14\tabcolsep) * \real{0.1250}}
  >{\raggedleft\arraybackslash}p{(\columnwidth - 14\tabcolsep) * \real{0.1250}}
  >{\raggedleft\arraybackslash}p{(\columnwidth - 14\tabcolsep) * \real{0.1250}}
  >{\raggedleft\arraybackslash}p{(\columnwidth - 14\tabcolsep) * \real{0.1250}}
  >{\raggedleft\arraybackslash}p{(\columnwidth - 14\tabcolsep) * \real{0.1250}}
  >{\raggedleft\arraybackslash}p{(\columnwidth - 14\tabcolsep) * \real{0.1250}}
  >{\raggedleft\arraybackslash}p{(\columnwidth - 14\tabcolsep) * \real{0.1250}}@{}}
\toprule\noalign{}
\begin{minipage}[b]{\linewidth}\raggedleft
\(\tau_0\)
\end{minipage} & \begin{minipage}[b]{\linewidth}\raggedleft
\(\beta\)
\end{minipage} & \begin{minipage}[b]{\linewidth}\raggedleft
\(n\)
\end{minipage} & \begin{minipage}[b]{\linewidth}\raggedleft
Admiss.
\end{minipage} & \begin{minipage}[b]{\linewidth}\raggedleft
Cov. \(\chi^2\)
\end{minipage} & \begin{minipage}[b]{\linewidth}\raggedleft
Cov. \(F\)
\end{minipage} & \begin{minipage}[b]{\linewidth}\raggedleft
Empty \(\chi^2\)
\end{minipage} & \begin{minipage}[b]{\linewidth}\raggedleft
Empty \(F\)
\end{minipage} \\
\midrule\noalign{}
\endhead
\bottomrule\noalign{}
\endlastfoot
0.2 & 1 & 2000 & 0.998 & 0.968 & 0.986 & 0.000 & 0.000 \\
0.2 & 1 & 8000 & 0.978 & 0.951 & 0.953 & 0.047 & 0.047 \\
0.2 & 2 & 2000 & 0.226 & 0.947 & 0.991 & 0.018 & 0.000 \\
0.2 & 2 & 8000 & 0.984 & 0.937 & 0.949 & 0.053 & 0.045 \\
0.2 & 3 & 2000 & 0.000 & -- & -- & -- & -- \\
0.2 & 3 & 8000 & 0.542 & 0.937 & 0.956 & 0.052 & 0.041 \\
0.5 & 1 & 2000 & 0.980 & 0.967 & 0.994 & 0.000 & 0.000 \\
0.5 & 1 & 8000 & 0.990 & 0.954 & 0.958 & 0.051 & 0.046 \\
0.5 & 2 & 2000 & 0.134 & 0.970 & 0.985 & 0.030 & 0.000 \\
0.5 & 2 & 8000 & 0.666 & 0.961 & 0.967 & 0.042 & 0.033 \\
0.5 & 3 & 2000 & 0.000 & -- & -- & -- & -- \\
0.5 & 3 & 8000 & 0.000 & -- & -- & -- & -- \\
\end{longtable}

The prespecified family-adequacy rule was triggered because \(G6\) was
selected in 64.6\% of admissible replications for
\((\tau_0,\beta,n)=(0.2,1,8000)\). The dated extension added
\(G7=(0.30,0.50,0.70,0.85)\) and rechecked it in all 6,000 replications.
It was selected in 128 of the 489 admissible replications in that
configuration, so \(P(G7\mid\mathrm{admissible})=0.262<0.5\), and
nowhere else. Table 2a reports the extended result, while Table 2b
displays the original benchmark and the extension side by side.

\textbf{Table 2b. Family-adequacy extension for
\((\tau_0,\beta,n)=(0.2,1,8000)\).}

\begin{longtable}[]{@{}
  >{\raggedright\arraybackslash}p{(\columnwidth - 10\tabcolsep) * \real{0.1667}}
  >{\raggedleft\arraybackslash}p{(\columnwidth - 10\tabcolsep) * \real{0.1667}}
  >{\raggedleft\arraybackslash}p{(\columnwidth - 10\tabcolsep) * \real{0.1667}}
  >{\raggedleft\arraybackslash}p{(\columnwidth - 10\tabcolsep) * \real{0.1667}}
  >{\raggedleft\arraybackslash}p{(\columnwidth - 10\tabcolsep) * \real{0.1667}}
  >{\raggedleft\arraybackslash}p{(\columnwidth - 10\tabcolsep) * \real{0.1667}}@{}}
\toprule\noalign{}
\begin{minipage}[b]{\linewidth}\raggedright
Family
\end{minipage} & \begin{minipage}[b]{\linewidth}\raggedleft
Median joint length
\end{minipage} & \begin{minipage}[b]{\linewidth}\raggedleft
Median ratio
\end{minipage} & \begin{minipage}[b]{\linewidth}\raggedleft
Terminal quantile \(q_{.9}\)
\end{minipage} & \begin{minipage}[b]{\linewidth}\raggedleft
Cov. \(\chi^2\)
\end{minipage} & \begin{minipage}[b]{\linewidth}\raggedleft
Cov. \(F\)
\end{minipage} \\
\midrule\noalign{}
\endhead
\bottomrule\noalign{}
\endlastfoot
\(G1\)--\(G6\) benchmark & 0.300 & 0.435 & 0.75 & 0.951 & 0.953 \\
\(G1\)--\(G7\) extension & 0.275 & 0.417 & 0.85 & 0.951 & 0.953 \\
\end{longtable}

The empty-set frequency changes only from 0.045 to 0.047. Thus the
extension lengthens the validated horizon without materially changing
inference on \(\tau\).

Admissibility, rather than conditional coverage, is the decisive result
in difficult designs. At \(n=2000\) and \(\beta=3\), no candidate grid
is usable. On the earliest grid, failures divide almost equally between
exact rank loss and largest-observation domination. Detailed audit shows
that several cell--cause combinations contain zero or one event at the
first node, producing exact sample singularity. Moving later alleviates
early cause sparsity but exhausts the worst cell: on \(G6\), the first
recorded failures are the minimum-survival screen (57.3\%) and the
minimum-risk-set screen (42.7\%), before tail diagnostics are evaluated.
Hence there is no admissible intermediate window at that sample size.

At \(n=8000\), the outcome depends on dependence and contrast. The
procedure admits a grid in 54.2\% of replications for \((0.2,3)\), but
in none for \((0.5,3)\); the latter fails primarily through
fourth-moment load. Conversely, low-contrast designs usually retain a
grid, though their survival confidence sets may remain broad. The
procedure therefore separates two limitations: weak information about
dependence and unstable cause-specific recovery.

Table 3 reports the selected horizon in pooled-duration quantiles and
separates information supplied by the survival block from information
supplied by the joint system. The two informativeness columns give
conditional probabilities, among admissible replications, that the
corresponding set occupies less than 65\% of the maintained
\(\tau\)-range. The last column reports the conditional probability that
the joint set is strictly shorter than the survival set.

\newpage

\textbf{Table 3. Selected horizons and the incremental information in
the cause indicator.}

\begin{longtable}[]{@{}
  >{\raggedleft\arraybackslash}p{(\columnwidth - 12\tabcolsep) * \real{0.1379}}
  >{\raggedleft\arraybackslash}p{(\columnwidth - 12\tabcolsep) * \real{0.1379}}
  >{\raggedleft\arraybackslash}p{(\columnwidth - 12\tabcolsep) * \real{0.1379}}
  >{\centering\arraybackslash}p{(\columnwidth - 12\tabcolsep) * \real{0.1724}}
  >{\raggedleft\arraybackslash}p{(\columnwidth - 12\tabcolsep) * \real{0.1379}}
  >{\raggedleft\arraybackslash}p{(\columnwidth - 12\tabcolsep) * \real{0.1379}}
  >{\raggedleft\arraybackslash}p{(\columnwidth - 12\tabcolsep) * \real{0.1379}}@{}}
\toprule\noalign{}
\begin{minipage}[b]{\linewidth}\raggedleft
\(\tau_0\)
\end{minipage} & \begin{minipage}[b]{\linewidth}\raggedleft
\(\beta\)
\end{minipage} & \begin{minipage}[b]{\linewidth}\raggedleft
\(n\)
\end{minipage} & \begin{minipage}[b]{\linewidth}\centering
Terminal quantile \(q_{.1}/q_{.5}/q_{.9}\)
\end{minipage} & \begin{minipage}[b]{\linewidth}\raggedleft
Survival inform.
\end{minipage} & \begin{minipage}[b]{\linewidth}\raggedleft
Joint inform.
\end{minipage} & \begin{minipage}[b]{\linewidth}\raggedleft
Joint narrows
\end{minipage} \\
\midrule\noalign{}
\endhead
\bottomrule\noalign{}
\endlastfoot
0.2 & 1 & 2000 & 0.40 / 0.55 / 0.55 & 0.044 & 0.106 & 0.427 \\
0.2 & 1 & 8000 & 0.55 / 0.75 / 0.85 & 0.358 & 0.998 & 0.998 \\
0.2 & 2 & 2000 & 0.20 / 0.30 / 0.40 & 0.080 & 0.221 & 0.239 \\
0.2 & 2 & 8000 & 0.65 / 0.65 / 0.75 & 1.000 & 1.000 & 0.935 \\
0.2 & 3 & 8000 & 0.40 / 0.55 / 0.55 & 1.000 & 1.000 & 0.469 \\
0.5 & 1 & 2000 & 0.30 / 0.40 / 0.40 & 0.127 & 0.273 & 0.369 \\
0.5 & 1 & 8000 & 0.40 / 0.40 / 0.55 & 0.471 & 0.994 & 1.000 \\
0.5 & 2 & 2000 & 0.20 / 0.30 / 0.40 & 0.537 & 0.642 & 0.254 \\
0.5 & 2 & 8000 & 0.20 / 0.30 / 0.40 & 1.000 & 1.000 & 0.901 \\
\end{longtable}

Configurations \((0.2,3,2000)\), \((0.5,3,2000)\), and \((0.5,3,8000)\)
are omitted because no replication produced an admissible cause-specific
grid, so the conditional quantities in this table are undefined.

The cause indicator is most consequential under the weakest contrast
considered. At \(n=8000\) and \(\beta=1\), the survival set is
informative in only 35.8\% and 47.1\% of admissible replications,
whereas the joint set is informative in 99.8\% and 99.4\%. This gain is
obtained despite increasing the 95\% reference critical value from
\(\chi^2_{4}=9.49\) to \(\chi^2_{16}=26.30\): uninformative additional
moments would tend to widen the inversion, so the observed near-halving
of length is conservative evidence that the cause-specific moments
contain substantial information about \(\tau\). The narrowing is not
purchased through undercoverage; joint chi-square coverage is 0.951 and
0.954 in the two \(\beta=1,n=8000\) designs.

At \(n=2000\), \(J\) narrows the set in only about one quarter to two
fifths of admissible replications. For \(\beta=2\) at \(n=8000\), the
survival set is already informative in every admissible replication and
\(J\) commonly shortens it further, but is not needed to cross the
prespecified informativeness threshold. At
\((\tau_0,\beta,n)=(0.2,3,8000)\), by contrast, the joint set is
strictly shorter in only 46.9\% of admissible replications and the
median length ratio equals one. Hence the large transition from weak
survival inference to informative joint inference is specific to the
weakest contrast examined. A lower boundary must exist as \(\beta\to0\),
where survival identification disappears, but it does not bind in the
simulated range; the binding limitation here is excessive contrast,
which makes one cause practically disappear in some cells.

\hypertarget{main-lessons}{%
\subsection{4.5 Main lessons}\label{main-lessons}}

First, categorical single crossing works without covariate derivatives,
but finite-sample precision is governed by contrast strength and the
maintained dependence range. Under the weakest contrast considered, the
cause indicator transforms a frequently uninformative survival inversion
into an informative joint inversion while preserving nominal coverage.
Second, the cause-specific block cannot be judged by nominal sample size
alone; event allocation within the worst cell and the influence-function
tail determine the usable horizon, and excessive contrast can destroy
admissibility even while strengthening the survival block. Third, empty
inverted sets behave as specification rejections rather than failed
point estimates. Finally, the operational output is a confidence set
together with its strength, boundary, admissibility, selected horizon,
rank, exposure, and tail diagnostics.

\hypertarget{scope-interpretation-and-extensions}{%
\section{5. Scope, interpretation, and
extensions}\label{scope-interpretation-and-extensions}}

The results establish that categorical exclusion restrictions in an
Archimedean competing-risks model have observable content. This
conclusion is stronger than identification alone but narrower than an
omnibus test of the latent competing-risks law. The maintained model
combines several components: the exclusion restrictions, the chosen
Archimedean copula family, stability of its dependence parameter across
time and covariate strata, and the substantive parameter range imposed
in Assumption 4. Rejection means that these components are jointly
incompatible with the observed law. Without additional structure, the
test cannot determine which component is responsible. Accordingly, a
nonempty inverted set should be interpreted as a set of model--parameter
pairs that remain observationally compatible, not as evidence that every
maintained assumption is literally true.

\hypertarget{conditional-exclusions-and-categorical-stratification}{%
\subsection{5.1 Conditional exclusions and categorical
stratification}\label{conditional-exclusions-and-categorical-stratification}}

An exclusion restriction need not hold marginally to be scientifically
defensible. Let \(X\) collect common observed covariates that may affect
both latent durations. The restrictions may instead be stated
conditionally as

\[
S_1(t\mid z_1,z_2,x)=S_1(t\mid z_1,x),\quad
S_2(t\mid z_1,z_2,x)=S_2(t\mid z_2,x).
\]

For categorical \(X\), the identification and testing arguments apply
within each stratum. This is an important practical distinction from
procedures based on covariate cross-derivatives: adding categorical
controls creates additional cells rather than a higher-dimensional
derivative-smoothing problem. It does not, however, make the exclusion
restrictions automatically valid. Unobserved common heterogeneity may
remain after conditioning, and the substantive justification for each
exclusion must be stated independently of the statistical test.

The price of conditioning is cell fragmentation. With two binary
excluded covariates and \(K\) strata of \(X\), the procedure uses \(4K\)
cells, and the effective information is governed by the least exposed
relevant cell rather than by the total sample size. Strata and any
pooling rule should therefore be specified before inspecting the
specification-test outcomes. Data-dependent merging of sparse cells
would otherwise introduce an additional selection layer not covered by
the fixed-design argument.

Pooling strata also imposes stability of the copula parameter. This is a
substantive restriction rather than a normalization. A useful diagnostic
is to construct stratum-specific survival-block confidence sets before
imposing a common \(\theta\). If these sets are mutually incompatible,
rejection of the pooled model should not be attributed to the exclusions
alone. If they overlap substantially, the common-parameter analysis
gains precision and supplies additional overidentifying restrictions
across strata.

\hypertarget{weak-categorical-variation}{%
\subsection{5.2 Weak categorical
variation}\label{weak-categorical-variation}}

The single-crossing result is qualitative: nonzero contrasts identify
the dependence parameter in the population. It does not imply that every
categorical design is informative in finite samples. The local
sensitivity of the cross-difference is driven by the product of the two
transformed survival contrasts. When either contrast is small, the
survival-block confidence set can occupy most of the maintained
parameter range even in a large sample. This is the categorical analogue
of a weak first stage.

For this reason, the point estimator is a secondary summary. The primary
inferential object is the inverted confidence set, accompanied by the
design- strength statistic \(\widehat{\mathcal S}\), its
boundary-contact indicators, and the fraction of the maintained range
that remains compatible with the data. In weak designs the procedure may
deliver correct coverage only because the confidence set is nearly
uninformative. Such a result should be reported as weak information, not
as successful precision. Conversely, an empty joint set represents
rejection of the maintained model and must not be replaced by the
nearest point estimate.

\hypertarget{the-data-supported-time-horizon}{%
\subsection{5.3 The data-supported time
horizon}\label{the-data-supported-time-horizon}}

Cause-specific recovery is necessarily local to the time interval over
which the moment system remains statistically usable. At a fixed time
point the weights are bounded whenever the worst-cell survival
probability is bounded away from zero, so standard finite-dimensional
asymptotics remain available. The finite-sample constants can
nevertheless deteriorate rapidly as the horizon moves into the tail. The
cause-specific influence functions combine the generator derivative with
the plug-in effect of estimating cell survival; late observations may
therefore dominate empirical second and fourth moments well before the
nominal risk set becomes empty.

The grid-selection rule makes this limitation explicit. It searches a
finite, prespecified family of candidate grids and retains the latest
grid satisfying the exposure, rank, fourth-moment, and
largest-observation diagnostics uniformly over the survival-block
confidence set. Failure to find an admissible grid is an output of the
procedure: dependence may still be partially informed by overall
survival while cause-specific recovery is withheld. The failure
annotation defined in Section 3 uses the earliest candidate grid to
distinguish inadequate worst-cell exposure from unstable
influence-function tails; it is descriptive and is evaluated only after
the whole candidate family has been checked.

The present exposure screens use cell survival and risk-set counts. They
do not guarantee adequate cause-specific event exposure in every cell.
Under strong covariate contrasts, an early pooled-time grid can lie in a
region where one cause is systematically almost absent from particular
cells even though their risk sets remain large. In the
\((\tau_0,\beta,n)=(0.2,3,2000)\) diagnostic, 82.3\% of the rank-failure
replications contained a zero cell--cause count at the first node and
99.6\% contained a count no larger than one. The smallest eigenvalue was
zero to machine precision, and the rank distribution was identical at
the first failure point and at \(\tau=0.5\). Thus the sample singularity
was systematic and parameter-independent in this configuration, not an
artifact of the numerical rank tolerance. This does not imply population
singularity or persistence as \(n\to\infty\) when the rare-event
probabilities are positive.

The confirmatory protocol reports this failure separately rather than
retroactively changing its admissibility rule. Future implementations
should add a prespecified conservative lower bound on expected
cause-specific event counts by cell and node, calibrated from the design
or an independent pilot sample and separated from its acceptance
threshold by a fixed margin, alongside the risk-set screens. Raw
observed counts remain useful diagnostics but should not become a
post-outcome selection rule.

This creates a two-sided horizon constraint at fixed sample size. Moving
the grid earlier can worsen cause-specific event sparsity even as the
risk sets grow, whereas moving it later depletes the worst-cell survival
and risk set and may also increase influence-function tail load. An
admissible intermediate window need not exist. In the diagnostic above,
\(G1\) failed through cause-specific rank loss, while on \(G6\) the
first binding conditions were \(\pi_{\min}\) in 57.3\% and \(Y_{\min}\)
in 42.7\% of the same replications.

The supported horizon should be reported together with the recovered
marginals. Extending the horizon generally requires a larger sample, but
there is no universal conversion from total sample size to usable
duration. The relevant constant depends on the dependence parameter,
cell imbalance, covariate contrasts, and the event-time grid. Thus the
pilot threshold for the effective sample size is a planning diagnostic,
not a universal regularity constant. The empirical curve mapping
terminal time into estimated effective size and largest-observation
share is more informative than a single nominal sample-size requirement.

\hypertarget{maintained-dependence-range-and-copula-family}{%
\subsection{5.4 Maintained dependence range and copula
family}\label{maintained-dependence-range-and-copula-family}}

Assumption 4 restricts Kendall's \(\tau\) to a compact, substantively
chosen range \(\mathcal T_0\). Coverage statements are conditional on
the true value belonging to this range. When an inverted set touches
\(\partial\mathcal T_0\), the corresponding endpoint is
assumption-censored. In particular, if the upper boundary binds, the
data identify only a lower confidence limit within the maintained range;
the numerical upper endpoint repeats the assumption rather than a
data-driven bound. This boundary is distinct from the edge of the wider
computational grid, contact with which indicates inadequate numerical
search.

Sensitivity to \(\mathcal T_0\) should be assessed using ranges chosen
before examining the inferential outcomes. Stability of the lower
endpoint when the upper maintained bound is expanded supports a
one-sided interpretation. If the lower endpoint moves materially, the
apparently data-driven part of the conclusion is itself sensitive to the
maintained range and the wider analysis should be primary.

All identification and testing statements are also conditional on the
copula family. A rejection cannot by itself separate failure of that
family from failure of the exclusions. A transparent robustness analysis
applies the procedure to a finite, prespecified collection of
one-parameter Archimedean families and reports the surviving pairs
\((m,\tau)\). The composite class is rejected only if every constituent
family is rejected. This intersection-- union rule requires no
multiplicity correction to control size under the union null. It does
not guarantee power: Clayton, Gumbel, and Frank copulas all approach
independence, so their observable implications must become difficult to
distinguish as \(\tau\) approaches zero. A continuously indexed
multi-parameter copula class would require a new identification
argument; scalar single crossing does not extend automatically to an
unordered parameter vector.

\hypertarget{right-censoring}{%
\subsection{5.5 Right censoring}\label{right-censoring}}

The present analysis is formulated for complete competing-risks
observations. This is the natural setting for the identification
question, because the classical nonidentifiability result already arises
when the first event time and cause are observed without censoring.
Independent right censoring does not change the population
cross-difference logic, but it changes estimation and the usable
horizon. Cell survival and cause-specific cumulative incidence can be
estimated by Kaplan--Meier and Aalen--Johansen methods, and their
martingale influence functions can be propagated through the same
finite-dimensional moment map.

This extension is not mechanically innocuous. Inverse censoring survival
enters the influence functions and multiplies the generator-induced tail
load. The admissible cause-specific horizon may therefore contract more
than suggested by event survival alone, especially in portfolios with
many recent entries and a common administrative extraction date: the
distribution of contract inception dates determines the censoring
survival \(G_c\), which can fall rapidly in a fast-growing portfolio.
The effective-size and largest-observation diagnostics must be
recomputed for the censored influence functions; values calibrated under
complete observation cannot be carried over. Developing and validating
this extension is left for subsequent work rather than appended without
a separate finite-sample analysis.

\hypertarget{broader-implication}{%
\subsection{5.6 Broader implication}\label{broader-implication}}

The broader question raised by the paper is not specific to Archimedean
copulas. In a nonidentified latent model, an identifying restriction may
either saturate the observable law or leave refutable implications. The
fixed-time dimension count provides useful bookkeeping: in the present
\(2\times2\) design, the overall-survival cross-difference supplies one
restriction, while three independent cause-specific contrasts restrict
the residual allocation. This count is only a necessary local check and
a way to locate the remaining freedom. It does not establish
identification, recoverability, or refutability. Recovery follows from
the constructive sufficiency argument in Theorem 1, while strict
refutability is established by the mass-redistribution counterexample.
Both results concern functional compatibility across time and cannot be
inferred from a count at one fixed time point.

A general characterization of when identifying restrictions possess
empirical content would extend beyond the present model. Such a theory
would need to distinguish local dimension counting from global
compatibility and to handle inequality, shape, and support restrictions.
The current competing-risks model provides a canonical example in which
the observable image, the remaining latent freedom, and the
specification statistic can all be written explicitly.

\hypertarget{appendix-a.-proofs}{%
\section{Appendix A. Proofs}\label{appendix-a.-proofs}}

\hypertarget{a.0-cause-sum-identity}{%
\subsection{A.0 Cause-sum identity}\label{a.0-cause-sum-identity}}

\textbf{Lemma A.0.} If \(\pi_c\) is continuous with \(\pi_c(0)=1\), then
for every candidate \(\theta\), \[
A_{1,\theta,c}(t)+A_{2,\theta,c}(t)=\phi_\theta\{\pi_c(t)\}.
\]

\textbf{Proof.} Let \(F_c=1-\pi_c\) be the distribution of the observed
minimum in cell \(c\). Because the two causes partition its events, \[
\begin{aligned}
A_{1,\theta,c}(t)+A_{2,\theta,c}(t)
&=\int_{(0,t]}-\phi_\theta'\{\pi_c(s)\}\,dF_c(s)\\
&=\int_{(0,t]}\phi_\theta'\{\pi_c(s)\}\,d\pi_c(s)\\
&=\phi_\theta\{\pi_c(t)\}-\phi_\theta\{\pi_c(0)\}
=\phi_\theta\{\pi_c(t)\},
\end{aligned}
\] where the last equality uses \(\pi_c(0)=1\) and \(\phi_\theta(1)=0\).
Specifically, continuity makes \(\pi_c\) a continuous function of
bounded variation, so the \(C^1\) chain rule for Lebesgue--Stieltjes
integrals gives \[
\phi_\theta\{\pi_c(t)\}-\phi_\theta\{\pi_c(0)\}
=\int_{(0,t]}\phi_\theta'\{\pi_c(s)\}\,d\pi_c(s).
\] \(\square\)

\hypertarget{a.1-proof-of-proposition-1}{%
\subsection{A.1 Proof of Proposition
1}\label{a.1-proof-of-proposition-1}}

At the true parameter, additive separability gives
\(\phi_{\theta_0}(\pi_{ab})=\alpha_a+\beta_b\). For a candidate
\(\theta\), write
\(H_{\theta,\theta_0}=\phi_\theta\circ\phi_{\theta_0}^{-1}\) and
\(s=\phi_{\theta_0}^{-1}(u)\). If \(\theta>\theta_0\), Assumption 2 says
that \[
r(s)=\frac{\phi_{\theta_0}'(s)}{\phi_\theta'(s)}
\] is strictly increasing in \(s\). Hence \[
H_{\theta,\theta_0}'(u)
=\frac{\phi_\theta'(s)}{\phi_{\theta_0}'(s)}
=\frac{1}{r(s)}
\] is strictly decreasing in \(s\). Since \(s=\phi_{\theta_0}^{-1}(u)\)
is strictly decreasing in \(u\), \(H_{\theta,\theta_0}'\) is strictly
increasing in \(u\), and \(H_{\theta,\theta_0}\) is strictly convex.
Reversing the parameter order shows that \(H_{\theta,\theta_0}'\) is
strictly decreasing, and \(H_{\theta,\theta_0}\) strictly concave, when
\(\theta<\theta_0\).

The mixed finite difference can be written without imposing a pointwise
sign on \(H''\):

\[
\Delta_t(\theta)=
\int_{\alpha_0}^{\alpha_1}
\{H_{\theta,\theta_0}'(u+\beta_1)
-H_{\theta,\theta_0}'(u+\beta_0)\}\,du.
\]

The integral is oriented when \(\alpha_1-\alpha_0<0\). Strict
monotonicity of \(H'\) makes the bracket have the sign of
\(\beta_1-\beta_0\) for \(\theta>\theta_0\) and the opposite sign for
\(\theta<\theta_0\). Therefore the integral has the sign stated in
\eqref{eq:single-crossing-sign}; when both contrasts are nonzero, it
vanishes only at \(\theta_0\). Thus one informative table identifies
\(\theta_0\). Nonzero contrasts are a design condition for this
proposition, not part of the generator-ordering assumption.

\hypertarget{a.2-proof-of-theorem-1}{%
\subsection{A.2 Proof of Theorem 1}\label{a.2-proof-of-theorem-1}}

Necessity follows directly from the model. Applying \(\phi_\theta\) to
the Archimedean survival representation and using the exclusions gives
additive separability and hence \eqref{eq:cross-difference}. The
copula-graphic identity \eqref{eq:copula-graphic} gives \[
A_{1,\theta,abx}=a_{a,x}+c_x,\qquad
A_{2,\theta,abx}=b_{b,x}-c_x
\] for a normalization \(c_x\) common to the four cells. Therefore the
residuals satisfy \eqref{eq:compatibility}. Because they are transforms
of survival functions, the two constructed transformed marginal
cumulative hazards start at zero, are nondecreasing, and remain in the
domain of \(\phi_\theta^{-1}\).

For sufficiency, suppose conditions (i)--(iii) of Theorem 1 hold. The
vanishing cross-difference permits an additive decomposition
\(\phi_\theta(\pi_{abx})=a_{a,x}+b_{b,x}\). By \eqref{eq:compatibility},
\(R_{ab,x}=c_x\), and \eqref{eq:cause-sum} therefore gives \[
A_{1,\theta,abx}=a_{a,x}+c_x,\qquad
A_{2,\theta,abx}=b_{b,x}-c_x.
\] Condition (iii) makes the functions in \eqref{eq:recovered-marginals}
valid survival functions. They obey the exclusions by construction.
Moreover, \[
\phi_\theta\{S_1(t\mid a,x)\}
+\phi_\theta\{S_2(t\mid b,x)\}
=a_{a,x}(t)+b_{b,x}(t)
=\phi_\theta\{\pi_{abx}(t)\},
\] so the Archimedean combination reproduces the observed overall
survival. It remains to verify the cause allocation without assuming
densities. Let \(\mu_{j,abx}=dF_{j,abx}\) denote the observed
cause-\(j\) subdistribution measure and put
\(w_\theta=-\phi_\theta'>0\). By the definition of the observable
transforms, \[
dA_{j,\theta,abx}(s)
=w_\theta\{\pi_{abx}(s)\}\,\mu_{j,abx}(ds).
\] The constructed marginals satisfy \[
d\phi_\theta\{S_1(s\mid a,x)\}=d\{a_{a,x}(s)+c_x(s)\}
=dA_{1,\theta,abx}(s)
\] and, analogously, \[
d\phi_\theta\{S_2(s\mid b,x)\}=d\{b_{b,x}(s)-c_x(s)\}
=dA_{2,\theta,abx}(s).
\] For an Archimedean survival law, its cause-\(j\) subdistribution
measure on the diagonal is \[
\widetilde\mu_{j,abx}(ds)
=\frac{d\phi_\theta\{S_j(s)\}}
{w_\theta\{\pi_{abx}(s)\}}.
\] Because Assumption 1 gives \(w_\theta>0\) throughout the retained
interior domain, division is valid. Substitution of the preceding
measure identities yields \(\widetilde\mu_{j,abx}=\mu_{j,abx}\) for
\(j=1,2\). Thus the constructed model reproduces the observed cause
allocation as well as the overall survival, proving sufficiency.

\hypertarget{a.3-proof-of-corollary-1}{%
\subsection{A.3 Proof of Corollary 1}\label{a.3-proof-of-corollary-1}}

It is enough to construct one observable law satisfying the survival
restriction but not the full model. Begin with a smooth interior law
satisfying the model at \(\theta_0\), with informative cell contrasts
and strictly positive cause-specific densities on a compact subinterval
\(I\subset(0,\bar t)\). In one cell \(c^\star\), choose a bounded
nonzero signed density \(h\), supported on \(I\), and define for
sufficiently small \(\varepsilon>0\) \[
dF^{\varepsilon}_{1,c^\star}
=dF_{1,c^\star}+\varepsilon h(s)\,ds,\qquad
dF^{\varepsilon}_{2,c^\star}
=dF_{2,c^\star}-\varepsilon h(s)\,ds.
\] Leave all other cells unchanged. Positivity of the baseline densities
on \(I\) ensures that both perturbed subdistributions remain nonnegative
for small enough \(\varepsilon\). Their sum is unchanged in every cell,
so the overall survival \(\pi\), and therefore every survival
cross-difference, is unchanged.

The perturbation changes the cause transform in cell \(c^\star\) by \[
\varepsilon\int_0^t -\phi_{\theta_0}'\{\pi_{c^\star}(s)\}h(s)\,ds
\] and changes the second transform by its negative. Choose \(h\) so
that this integral is nonzero for some retained \(t\). Then the four
residuals cannot remain equal, and \eqref{eq:compatibility} fails at
\(\theta_0\). Proposition 1 makes \(\theta_0\) the unique parameter
satisfying the unchanged survival restriction. Consequently no candidate
parameter rationalizes the perturbed observable law, even though it
remains in the survival-restricted class. The inclusion is therefore
strict.

\hypertarget{references}{%
\section{References}\label{references}}

Hansen, L. P., Heaton, J., and Yaron, A. (1996). Finite-sample
properties of some alternative GMM estimators. \emph{Journal of Business
and Economic Statistics}, 14(3), 262--280.

Heckman, J. J., and Honoré, B. E. (1989). The identifiability of the
competing risks model. \emph{Biometrika}, 76(2), 325--330.

Hiabu, M., Lo, S. M. S., and Wilke, R. A. (2025). Identifiability and
estimation of the competing risks model under exclusion restrictions.
\emph{Statistica Neerlandica}, 79(1), e70003.
https://doi.org/10.1111/stan.70003.

Liu, R. (2020). A competing risks model with time-varying heterogeneity
and simultaneous failure. \emph{Quantitative Economics}, 11(2),
535--577. https://doi.org/10.3982/QE1159.

Mourifié, I., Henry, M., and Méango, R. (2020). Sharp bounds and
testability of a Roy model of STEM major choices. \emph{Journal of
Political Economy}, 128(8), 3220--3283. https://doi.org/10.1086/708724.

Rivest, L.-P., and Wells, M. T. (2001). A martingale approach to the
copula-graphic estimator for the survival function under dependent
censoring. \emph{Journal of Multivariate Analysis}, 79(1), 138--155.

Stock, J. H., and Wright, J. H. (2000). GMM with weak identification.
\emph{Econometrica}, 68(5), 1055--1096.

Tsiatis, A. (1975). A nonidentifiability aspect of the problem of
competing risks. \emph{Proceedings of the National Academy of Sciences},
72(1), 20--22.

Zheng, M., and Klein, J. P. (1995). Estimates of marginal survival for
dependent competing risks based on an assumed copula. \emph{Biometrika},
82(1), 127--138.

\end{document}